\documentclass[preprint,prb]{revtex4}

\usepackage{amsmath}

\newcommand{\bfv}{\boldsymbol{v}}

\newcommand{\gradv}{\boldsymbol{\nabla}}
\def\v#1{{\bf#1}}

\begin{document}

\title{The Galilean limits of Maxwell's equations}
\author{Jos\'e A. Heras}
\email{herasgomez@gmail.com}
\affiliation{Universidad Aut\'onoma Metropolitana Unidad Azcapotzalco, Av. San Pablo No. 180, Col. Reynosa, 02200, M\'exico D. F. M\'exico}

\begin{abstract}
We show that if Maxwell's equations are expressed in a form independent of specific units, at least three Galilean limits can be extracted. The electric and magnetic limits can be regarded as nonrelativistic limits because they are obtained using the condition $|\bfv| \ll c$ and restrictions on the magnitudes of the sources and fields. The third limit is called the instantaneous limit and is introduced by letting $c\to \infty$. The electric and instantaneous limits have the same form, but their interpretation is different because the instantaneous limit cannot be considered as a nonrelativistic limit. We emphasize the double role that the speed of light $c$ plays in Maxwell's equations.
\end{abstract}

\maketitle

\section{INTRODUCTION}

The nonrelativistic behavior of Maxwell's equations is not usually discussed in textbooks on electromagnetic theory and special relativity.
This omission is surprising because these textbooks usually emphasize the relativistic invariance of Maxwell's equations. One possible explanation for this omission is that the nonrelativistic behavior of relativistic expressions can be subtle, as is the case for the Lorentz transformations for which there are two nonrelativistic limits.\cite{1,2} Consider the Lorentz transformations:
\begin{equation}
\label{1}
\v x' = \v x - \gamma \frac{\bfv}{c} ct + (\gamma - 1)\frac{\bfv(\bfv\cdot\v x)}{\bfv^2}, \quad
ct' = \gamma\bigg[ct - \frac{\bfv}{c} \cdot \v x\bigg],
\end{equation}
where $\gamma=(1-\bfv^2/c^2)^{-1/2}$, $\bfv$ is the relative velocity between two inertial (primed and unprimed) frames, and $c$ is the speed of light in vacuum (or equivalently, $c$ is the speed of propagation of the electric and magnetic fields in vacuum). If we assume that $|\bfv| \ll c$, it follows that $\gamma\to 1$. If, in addition, we assume the ultra-timelike condition\cite{2} $ct \gg |\v x|$, then Eq.~(\ref{1}) reduces to the Galilean transformations,
\begin{equation}
\label{2}
\v x'=\v x- \bfv t,\quad t'=t.
\end{equation}
which are the nonrelativistic ultra-timelike limit of the Lorentz transformations.

We can also have the ultra-spacelike condition\cite{2} $ct \ll |\v x|$, which can be used with $|\bfv| \ll c$ in Eq.~(\ref{1}) to obtain the nonrelativistic ultra-spacelike limit of the Lorentz transformations:
\begin{equation}
\label{3}
\v x'=\v x, \quad t'=t-\frac{1}{c^2} \bfv\cdot \v x.
\end{equation}
These transformations are known as ``Carroll transformations" in honor of the author of Alice in Wonderland.\cite{3} The reason for this name is that some predictions of Eq.~(\ref{3}) do not seem to be physically realizable. In particular, causal relations between events seem to be impossible.\cite{3}

Equation~(\ref{1}) admits a third limit which may be called the instantaneous limit. By letting $c\to \infty$ and then $\gamma=1$ in Eq.~(\ref{1}) expressed as $\v x' = \v x - \gamma \bfv , t + (\gamma - 1)[\bfv(\bfv\cdot\v x)]/\bfv^2$ and $t' = \gamma[t - \bfv\cdot\v x/c^2]$, we again obtain the Galilean transformations
\begin{equation}
\label{4}
\v x'=\v x- \bfv t,\quad t'=t.
\end{equation}
The effect of $c\to \infty$ consists in changing the finite speed of propagation of the electric and magnetic fields to an infinite speed of propagation, or equivalently, $c\to \infty$ transforms retarded fields into instantaneous fields.\cite{4} We note that the idea of instantaneous fields is natural in the context of a Galilei-invariant theory.

The instantaneous limit in Eq.~(\ref{4}) and the nonrelativistic ultra-timelike limit in Eq.~(\ref{2}) have the same form. However, Eqs.~(\ref{2}) and (\ref{4}) have a different interpretation.\cite{5} Equation~(\ref{2}) is valid when $|\bfv| \ll c$ and $ct \gg |\v x|$. These conditions are not required for the validity of Eq.~(\ref{4}). The ultra-timelike limit requires that $c$ be a fixed number in the chosen units, and the instantaneous limit arises when $c$ is made infinite. The instantaneous limit cannot generally be considered as a nonrelativistic limit because it allows finite but otherwise arbitrary velocities. When we take the limit $c\to \infty$ to $|\bfv|/c$ with $|\bfv|$ finite but otherwise arbitrary, then we formally obtain the result $\lim_{c\to\infty}|\bfv|/c=0$, which is independent of the nonzero value of $|\bfv|$.

This way of obtaining the two nonrelativistic limits of the Lorentz transformations can be applied to electromagnetic quantities.\cite{2} This procedure involves the low-velocity limit $|\bfv| \ll c$ (keeping $c$ as fixed) and assumes restrictions on the magnitudes of the relevant quantities. In electromagnetism with sources in free space there are two fields ($\v E$ and $\v B$) and two sources ($\rho$ and $\v J$). Restrictions on the magnitudes of these quantities can be assumed (in addition to $|\bfv| \ll c$) such that two nonrelativistic limits for electromagnetic quantities are obtained. The existence of two nonrelativistic limits for Maxwell's equations is then an expected result.

Le Bellac and L\'evy-Leblond\cite{1} obtained these two limits, which are known as the electric and magnetic limits of Maxwell's equations. They initiated the tradition of using SI units to derive the electric and magnetic limits. They wrote ``Any system, such as the CGS one, where $c$ enters in the very definition of units is bound to give inconsistent results."\cite{1} The existence of the electric and magnetic limits of Maxwell's equations has remained undisputed for more than thirty years. Classical and quantum aspects of these limits have been discussed in Refs.~\onlinecite{2,6}.
Nevertheless, a subtle and important question has remained. Why has the idea of two nonrelativistic limits for Maxwell's equations only been developed in SI units? No physical result should depend on the choice of a specific system of units. In particular, we should be able to obtain the electric and magnetic limits of Maxwell's equations in Gaussian units.

The procedure for obtaining the instantaneous limit of the Lorentz transformations can be applied to electromagnetic quantities by letting $c\to \infty$. Therefore an instantaneous limit for Maxwell's equations is expected, which would allow finite but otherwise arbitrary velocities. However, in letting $c\to \infty$ we must proceed with care because the parameter $c$ in Maxwell's equations plays a double role. On one hand, $c$ is a fixed quantity in a specific system of units and should not be modified. On the other hand, $c$ also represents the speed of propagation of the electromagnetic fields in any system of units, and thus it should be modified if we want to change the speed of propagation. To obtain the instantaneous limit of Maxwell's equations we must keep intact the $c$ associated with the system of units and change the $c$ of the propagation of fields. The question naturally arises: How do we distinguish the $c$ associated with units from the $c$ associated with propagation?

We discuss the dual role of $c$ in Maxwell's equations in Sec.~II and introduce the $c$ equivalence principle, which roughly says that the $c$ of units is equivalent to the $c$ of propagation.
We then express Maxwell's equations in a form independent of specific units,\cite{7} which involves the parameters $\alpha$, $\beta$, and $c$. This form makes transparent the double role of $c$ and is appropriate for obtaining the nonrelativistic and instantaneous limits of electromagnetic quantities. We show in Secs.~III and IV that Maxwell's equations in this form have three Galilean limits. The first two limits are the nonrelativistic electric and magnetic limits as introduced (in SI units) in Ref.~\onlinecite{1}. The third limit is the instantaneous limit, which is introduced by letting $c\to \infty$. The electric and instantaneous limits have the same form but a different physical interpretation. We discuss an alternative approach to deriving the electric and magnetic limits in Sec.~V and discuss the Lorentz force in the context of these two limits in Sec.~VI. We present our conclusions in Sec.~VII.

\section{The two roles of $C$ and The $C$ equivalence principle}

To discuss the first role of $c$, which we denote by $c_u$, consider the magnitude of the force between two equal charges $q$ separated by the distance $R$:
\begin{equation}
\label{5}
F=\bigg(\frac{\alpha}{4\pi}\bigg)\frac{q^2}{R^2},
\end{equation}
where the constant $\alpha$ is determined by the choice of units and the factor $4\pi$ is introduced for convenience. Consider the magnitude of the force per unit length between two infinitely long and parallel currents $I$ separated by $R$:
\begin{equation}
\label{6}
\frac{dF}{dl}=\bigg(\frac{\beta\chi}{4\pi}\bigg)\frac{2I^2}{R},
\end{equation}
where $\beta$ and $\chi$ are constants determined by the units ($\beta$ here does not represent $|\v v|/c$). From Eqs.~(\ref{5}) and (\ref{6}) we see that $\alpha q^2/R^2$ and $\beta\chi I^2$ must have the same dimensions because the associated forces must have the same dimensions. From this relation and the fact that the dimensions of $I$ are $q/T$ we conclude that the ratio $\alpha/(\beta\chi)$ has the dimensions of a velocity squared. What is this mysterious velocity? Before trying to answer this question, we point out that the unit of charge can be defined by specifying either $\alpha$ or $\beta\chi$. If either $\alpha$ or $\beta\chi$ is specified, then the value of the other quantity must be determined experimentally. In SI units we chose $\beta=\mu_0= 4\pi\times 10^{-7}$\,N/A$^2$ and $\chi=1$ and experimentally obtain\cite{8} $\alpha=1/\epsilon_0$ with $\epsilon_0= 8.85\times 10^{-12}$\,N/m. With these SI values of $\alpha$, $\beta$, and $\chi$ we obtain $\alpha/(\beta\chi)=1/(\epsilon_0\mu_0)$ or
\begin{equation}
\label{7}
\frac{\alpha}{\beta\chi}=c_u^2,
\end{equation}
where $c_u=2.9986\times 10^5$\,km/s. This value of $c_u$ tempts us to identify $c_u$ with the speed of propagation of light $c$ in vacuum. Before accepting this identification we need to consider other units. For Gaussian units we let $\alpha=4\pi$ and experimentally obtain $\beta\chi=4\pi/c_u^2$. For these units we chose $\beta=4\pi/c_u$ and $\chi=1/c_u$. For Heaviside-Lorentz units we let $\alpha=1$ and obtain $\beta\chi=1/c_u^2$. In this case we chose $\beta=1/c_u$ and $\chi=1/c_u$. In Table~I we collect the values of $\alpha, \beta$, and $\chi$ for Gaussian, SI, and Heaviside-Lorentz units.\cite{7}

All of these values for $\alpha, \beta$ and $\chi$ satisfy Eq.~(\ref{7}), which means that $c_u^2$ in Eq.~(\ref{7}) is independent of the choice of units and $c_u$ may be considered to be a fundamental constant of nature. We emphasize that $c_u$ results from action-at-a-distance laws [Eqs.~(\ref{5}) and (\ref{6})] where an instantaneous propagation is assumed, or equivalently, where the speed of propagation is taken to be infinity. Therefore the identification of $c_u$ with the speed of propagation of light $c$ (which is determined by optical means) is not evident from a physical point of view.

With the aid of the parameters $\alpha$ and $\beta$ we can express the static limit of Maxwell's equations in a form independent of units: $\gradv\cdot\v E=\alpha\rho,\gradv\times \v E= 0$, $\gradv\cdot\v B= 0$, and $\gradv\times \v B=\beta\v J$. This static limit does not contain the speed $c$ because there is no propagation speed for fields in the static regime.

The second role that $c$ plays in Maxwell's equations is the propagation speed of the fields $\v E$ and $\v B$. We will keep the letter $c$ to identify this second role, which we discuss using a heuristic argument starting with a time-dependent generalization of the static Maxwell's equations: $\gradv\cdot\v E(\v x,t)=\alpha\rho(\v x,t),\gradv\times \v E(\v x,t)= 0$, $\gradv\cdot\v B(\v x,t)= 0 $, and $\gradv\times \v B(\v x,t)=\beta\v J(\v x,t)$. These equations define an instantaneous action-at-a-distance theory but are inconsistent with the continuity equation. This objection can be overcome by adding the term $k\partial\v E(\v x,t)/\partial t$ to the equation $\gradv\times \v B(\v x,t)=\beta\v J(\v x,t)$, where $k$ is a constant to be determined. Then we have $\gradv\times \v B(\v x,t)+k\partial \v E(\v x,t)/\partial t=\beta\v J(\v x,t)$. We take the divergence of this equation and use $\gradv\cdot\v E(\v x,t)=\alpha\rho(\v x,t)$ and obtain $[-\beta/(k\alpha)]\gradv\cdot\v J(\v x,t) +\partial\rho(\v x,t)/\partial t =0$, which becomes the continuity equation when $k=-\beta/\alpha$. Therefore the modified equations are
\begin{subequations}
\label{8}
\begin{align}
\gradv\cdot\v E&=\alpha\rho \label{8a}\\
\gradv\times \v E& = 0, \label{8b}\\
\gradv\cdot\v B&= 0, \label{8c}\\
\gradv\times \v B-\frac{\beta}{\alpha}\frac{\partial \v E}{\partial t}& =\beta\v J. \label{8d}
\end{align}
\end{subequations}
These imply the continuity equation but they still represent an instantaneous action-at-a-distance theory where the fields $\v E$ and $\v B$ propagate instantaneously. A further modification of Eq.~(\ref{8}) is required to obtain a field theory with retarded fields. We take the curl of Eq.~(\ref{8d}) and use Eq.~(\ref{8c}) to obtain
\begin{equation}
\label{9}
\nabla^2\v B +\frac{\beta}{\alpha}\frac{\partial}{\partial t}\gradv\times\v E=-\beta\gradv\times\v J.
\end{equation}
Equation~\eqref{9} coincides with the wave equation for the field $\v B$:
\begin{equation}
\label{10}
\nabla^2\v B -\frac{1}{c^2}\frac{\partial^2\v B}{\partial t^2}=-\beta\gradv\times\v J,
\end{equation}
where $c$ is the propagation speed of the field $\v B$ if we modify Eq.~(\ref{8b}) as
\begin{equation}
\label{11}
\gradv\times \v E=-\frac{\alpha}{\beta c^2}\frac{\partial \v B}{\partial t}.
\end{equation}
With Eq.~(\ref{11}) we obtain Maxwell's equations expressed in a way independent of units:
\begin{subequations}
\label{12}
\begin{align}
\gradv\cdot\v E&=\alpha\rho,\label{12a}\\
\gradv\times \v E+\frac{\alpha}{\beta c^2}\frac{\partial \v B}{\partial t}&= 0, \label{12b}\\
\gradv\cdot\v B&= 0, \label{12c}\\
\gradv\times \v B-\frac{\beta}{\alpha}\frac{\partial \v E}{\partial t}&=\beta\v J. \label{12d}
\end{align}
\end{subequations}
Note the distinctive role that Faraday's law plays in Maxwell's equations: It is the only one of the equations that contains the speed of propagation $c$. The other equations implicitly contain the velocity $c_u$ inside the parameters $\alpha$ and $\beta$. The wave equations for $\v B$ and $\v E$ are given by Eq.~(\ref{10}) and
\begin{equation}
\label{13}
\nabla^2\v E -\frac{1}{c^2}\frac{\partial^2 \v E}{\partial t^2} =\alpha \gradv\rho+\frac{\alpha}{c^2}\frac{\partial \v J}{\partial t}.
\end{equation}
The retarded solutions of Eqs.~(\ref{10}) and (\ref{13}) show that $c$ is the speed of propagation of the fields $\v E$ and $\v B$.
From Eq.~(\ref{7}) we obtain the relation $\alpha/\beta=\chi c_u^2$, which can be used in Eq.~(\ref{12b}) to obtain the alternative form:
\begin{equation}
\label{14}
\gradv\times \v E+\chi\frac{c_u^2}{c^2}\frac{\partial \v B}{\partial t}= 0,
\end{equation}
which explicitly displays the speeds $c_u$ and $c$. In SI units $(\chi=1)$ we have
\begin{equation}
\label{15}
\gradv\times \v E+\frac{c_u^2}{c^2}\frac{\partial \v B}{\partial t}= 0.
\end{equation}
The experimentally verified existence of electromagnetic waves propagating with the speed $c=2.9986\times 10^5$\,km/s leads to the numerical equality
\begin{equation}
\label{16}
c_u=c,
\end{equation}
which allows us to write Eq.~(\ref{15}) in its well-known SI form:
\begin{equation}
\label{17}
\gradv\times \v E+\frac{\partial \v B}{\partial t}= 0.
\end{equation}

Numerical coincidence does not necessarily mean physical coincidence. The speed $c_u$ was obtained by introducing units in the context of instantaneous action-at-a-distance forces (with no radiation), and the speed $c$ was obtained by changing the instantaneous propagation of the fields to a finite propagation to obtain a field interaction (with radiation).

To some extent the situation is similar to the equality of the inertial mass $m_i$ of a body and its gravitational mass $m_g$. The inertial mass and the gravitational mass represent different physical properties, but repeated experiments have demonstrated that both masses are numerically equivalent: $m_i=m_g$. By the equivalence principle $m_g/m_i=1$. By analogy, we can say that Eq.~(\ref{16}) is the mathematical representation of the $c$ equivalence principle, which can be stated as follows: The speed $c_u$ obtained in the process of defining electromagnetic units via action-at-a-distance forces is equivalent to the propagation speed $c$ of electromagnetic waves in vacuum. In short, the $c_u$ of action-at-a-distance is equivalent to the $c$ of field action.
This principle formalizes the empirical equality in Eq.~(\ref{16}).

To illustrate this principle, consider the electric field $\v E$ in terms of the potentials $\Phi$ and $\v A$. From Eq.~(\ref{12}) we obtain $\v E=-\gradv\Phi-\chi(c_u^2/c^2)\partial \v A/\partial t$, which depends on the ratio $c_u/c$. By the $c$ equivalence principle we have $c_u/c=1$, and hence we obtain $\v E=-\gradv\Phi-\chi \partial \v A/\partial t$, which becomes $\v E=-\gradv\Phi-\partial \v A/\partial t$ in SI units.

We can also use the Ampere-Maxwell law in Eq.~(\ref{12}) and Faraday's law in Eq.~(\ref{14}) to obtain the energy density associated with the electric and magnetic fields: $u=(1/2)[\v E^2/\alpha +(c_u^2/c^2)\chi\v B^2/\beta]$, which also depends on $c_u/c$. By invoking the $c$ equivalence principle we have $c_u/c=1$, and therefore $u=(1/2)[\v E^2/\alpha + \chi\v B^2/\beta]$, which becomes the familiar expression $u=(1/2)[\epsilon_0\v E^2 + \v B^2/\mu_0]$ in SI units.

\section{Electric and magnetic limits}

Consider the relativistic transformations of the electric and magnetic fields and the charge and current densities:
\begin{subequations}
\label{18}
\begin{align}
\v E' &= \gamma\bigg[\v E + \frac{\bfv}{c} \times \frac{\alpha}{\beta c}\v B\bigg] + (\gamma - 1)\frac{\bfv(\bfv \cdot \v E)}{\bfv^2},\\
\frac{\alpha}{\beta c}\v B'&= \gamma\bigg[\frac{\alpha}{\beta c}\v B - \frac{\bfv}{c} \times \v E\bigg] + (\gamma - 1)\frac{\bfv[\bfv \cdot \alpha\v B/(\beta c)]}{\bfv^2},\\
c\rho' &= \gamma\bigg[c\rho - \frac{\bfv}{c} \cdot \v J\bigg], \quad \v J' = \v J - \gamma\frac{\bfv}{c} c\rho + (\gamma - 1)\frac{\bfv(\bfv \cdot \v J)}{\bfv^2}.
\end{align}
\end{subequations}
We now define the electric and magnetic limits.

In {\it electric limit} the field $\v E$ is dominant with respect to the field $\alpha\v B/(\beta c)$ and the density $c\rho$ is dominant with respect to the current density $\v J$:
\begin{equation}
\label{19}
|\v E| \gg \frac{\alpha}{\beta c}|\v B|,\quad c|\rho| \gg |\v J|.
\end{equation}
We use Eq.~(\ref{19}) and $|\bfv| \ll c$ (and then $\gamma\to 1$) so that Eq.~(\ref{18}) reduces to
\begin{subequations}
\label{20}
\begin{align}
\v E'&=\v E,\label{20a}\\
\v B'&=\v B-\frac{\beta}{\alpha}\bfv\times \v E, \label{20b}\\
\rho'&=\rho, \label{20c}\\
\v J' &= \v J -\bfv\rho.\label{20d}
\end{align}
\end{subequations}
From Eq.~(\ref{2}) we can derive the Galilean transformations:
\begin{equation}
\label{21}
\gradv'=\gradv, \quad \frac{\partial}{\partial t'}=\frac{\partial}{\partial t}+\bfv\cdot\gradv.
\end{equation}
Although Eq.~(\ref{12}) is not invariant under Eqs.~(\ref{20}) and (\ref{21}), a modification of Eq.~(\ref{12}) turns out to be invariant under Eqs.~(\ref{20}) and (\ref{21}). We define the electric limit of Maxwell's equations as the modification of Eq.~(\ref{12}) that remains invariant under Eqs.~(\ref{20}) and (\ref{21}). Our task is to identify the particular law in Eq.~(\ref{12}) that must be modified or replaced to define the electric limit. A direct calculation shows that $\gradv\cdot\v E=\alpha\rho$ and $\gradv\times \v B-(\beta/\alpha)\partial \v E/\partial t=\beta\v J$ are invariant under Eqs.~(\ref{20}) and (\ref{21}) [see Eqs.~(\ref{24a}) and (\ref{24i})]. However, the following result shows that $\gradv\times\v E+ [\alpha/(\beta c^2)]\partial \v B/\partial t= 0$ is not invariant under Eqs.~(\ref{20}) and (\ref{21}):
\begin{eqnarray}
\label{22}
\gradv' \times \v E' + \frac{\alpha}{\beta c^2}\frac{\partial \v B'}{\partial t'} &= & \gradv \times \v E + \frac{\alpha}{\beta c^2}\frac{\partial \v B}{\partial t} \nonumber \\
&&{} + \frac{\alpha}{\beta c^2}(\bfv \cdot \gradv)\v B - \frac{1}{c^2}\bfv \times \frac{\partial \v E}{\partial t} - \frac{1}{c^2}(\bfv \cdot \gradv)\bfv \times \v E.
\end{eqnarray}
The sum of the last three terms is nonzero in general. It follows that at least Faraday's law must be modified in the electric limit. To see how this modification must be done, consider $\gradv\cdot \v B=0$. By using Eqs.~(\ref{20}) and (\ref{21}) we obtain the relation [see Eq.~(24d)]: $\gradv'\cdot\v B'=\gradv\cdot\v B+(\beta/\alpha)\bfv\cdot(\gradv\times\v E)$. From this relation it follows that $\gradv\cdot \v B=0$ is not invariant unless we assume that $\gradv\times\v E=0$, which is invariant because $\gradv'=\gradv$ and $\v E'=\v E$ in the electric limit. These results indicate that Faraday's law $\gradv\times\v E+ [\alpha/(\beta c^2)]\partial \v B/\partial t= 0$ must be replaced by $\gradv\times\v E = 0$ in the electric limit. Therefore, the electric limit of Maxwell's equations is defined by
\begin{subequations}
\label{23}
\begin{align}
\gradv\cdot\v E&=\alpha\rho, \label{23a}\\
\gradv\times \v E& = 0,\label{23b}\\
\gradv\cdot\v B&= 0, \label{23c}\\
\gradv\times \v B-\frac{\beta}{\alpha}\frac{\partial \v E}{\partial t}
&=\beta\v J. \label{23d}
\end{align}
\end{subequations}
The truncated Faraday law, $\gradv\times\v E = 0$, should not be interpreted as the result of making $\partial \v B/\partial t= 0$ in the complete Faraday's law $\gradv\times\v E+ [\alpha/(\beta c^2)]\partial \v B/\partial t= 0$, but as the result of assuming conditions that imply that the term $[\alpha/(\beta c^2)]|\partial\v B/\partial t|$ is small with respect to $|\gradv\times \v E|$.\cite{9} In Sec.~V we will discuss such conditions.

To verify the Galilean invariance of Eq.~(\ref{23}) we use Eqs.~(\ref{20}), (\ref{21}), and (\ref{23}) and obtain
\begin{subequations}
\label{24}
\begin{align}
\gradv'\cdot\v E'-\alpha\rho'& = \gradv\cdot\v E-\alpha\rho,\label{24a}\\
\gradv'\cdot\v B' & = \gradv\cdot \Big[\v B-\frac{\beta}{\alpha} \bfv\times\v E \Big]\label{24b} \\
& = \gradv\cdot\v B-\frac{\beta}{\alpha}\gradv\cdot(\bfv\times\v E)\label{24c}\\
& = \gradv\cdot\v B+\frac{\beta}{\alpha} \bfv\cdot(\gradv\times\v E)\label{24d} \\
&= \gradv\cdot\v B,\label{24e} \\
\gradv'\times \v E'&=\gradv\times \v E,\label{24f}\\
\gradv' \times \v B' - \frac{\beta}{\alpha}\frac{\partial \v E'}{\partial t'} - \beta\v J'& =
 \gradv \times \Big[\v B-\frac{\beta}{\alpha}\bfv\times\v E \Big]
- \frac{\beta}{\alpha} \Big[\frac{\partial}{\partial t}
 + \bfv \cdot \gradv \Big]\v E - \beta[\v J -\bfv\rho]\label{24g} \\
& =\gradv \times \v B - \frac{\beta}{\alpha}\frac{\partial \v E}{\partial t} - \beta\v J
- \frac{\beta}{\alpha} \Big[ \gradv \times (\bfv \times \v E) + (\bfv \cdot \gradv)\v E - \bfv(\gradv\cdot\v E) \Big]\label{24h}\\
&= \gradv \times \v B - \frac{\beta}{\alpha}\frac{\partial \v E}{\partial t} - \beta\v J,\label{24i}
\end{align}
\end{subequations}
where we have used the identities
\begin{subequations}
\label{25}
\begin{align}
\gradv\cdot(\v a\times\v b) &= \v b\cdot(\gradv\times\v a)-\v a\cdot(\gradv\times\v b),\\
\gradv\times(\v a\times\v b) &= \v a(\gradv\cdot\v b) - \v b(\gradv\cdot\v a) + (\v b\cdot\gradv)\v a - (\v a\cdot\gradv)\v b.
\end{align}
\end{subequations}
The Galilean invariance of Eq.~(\ref{23}) directly follows from Eq.~(\ref{24}).\cite{10}

The electric limit describes quasistatic situations where electric effects are dominant. The equations in the electric limit correspond to those of the electroquasistatics approximation,\cite{11}  which describes capacitive but not inductive effects.\cite{12} We note that the energy density in the electric limit $u=\v E^2/(2\alpha)$ is associated only with the electric field. We also note that electrohydrodynamics relies on the electric limit.\cite{13}

In the {\it magnetic limit} the field $\alpha\v B/(\beta c)$ is dominant with regard to the field $\v E$ and also the current density $\v J$ is dominant with regard to density $c\rho$:
\begin{equation}
\label{26}
|\v E| \ll \frac{\alpha}{\beta c}|\v B|,\quad c|\rho| \ll |\v J|.
\end{equation}
We use Eq.~(\ref{26}) and $|\bfv| \ll c$ so that Eq.~(\ref{18}) reduces to
\begin{subequations}
\label{27}
\begin{align}
\v E'&=\v E +\frac{\alpha}{\beta c^2} \bfv\times \v B, \\
 \v B'&=\v B,\\
\rho'&=\rho-\frac{1}{c^2} \bfv\cdot\v J,\\
\v J'&=\v J.
\end{align}
\end{subequations}
Again, Eq.~(\ref{12}) is not invariant under Eqs.~(\ref{21}) and (\ref{27}). We define the magnetic limit of Maxwell's equations as the modification of Eq.~(\ref{12}) that remains invariant under Eqs.~(\ref{21}) and (\ref{27}). We can show that $\gradv\cdot\v B=0$ and $\gradv\times\v E+ [\alpha/(\beta c^2)]\partial \v B/\partial t= 0$ are invariant under Eqs.~(\ref{21}) and (\ref{27}) [see Eqs.~(\ref{30f}) and (\ref{30i})]. But the equation $\gradv\times \v B-(\beta/\alpha)\partial \v E/\partial t=\beta\v J$ is not invariant under Eqs.~(\ref{21}) and (\ref{27}):
\begin{equation}
\label{28}
\gradv' \times \v B' - \frac{\beta}{\alpha}\frac{\partial \v E'}{\partial t'} - \beta\v J' = \gradv \times \v B - \frac{\beta}{\alpha}\frac{\partial \v E}{\partial t} - \beta\v J - \frac{\beta}{\alpha}(\bfv\cdot \gradv)\v E - \frac{1}{c^2} \bfv \times \frac{\partial \v B}{\partial t} - \frac{1}{c^2}(\bfv \cdot \gradv)\bfv \times \v B.
\end{equation}
The sum of the last three terms is nonzero in general. It follows that the Ampere-Maxwell law must be modified or replaced in the magnetic limit. We can show that $\gradv' \cdot \v E'-\alpha\rho' = \gradv \cdot \v E-\alpha\rho-[\alpha/(\beta c^2)]\bfv\cdot[\gradv \times\v B-\beta\v J]$ [see Eq.~(\ref{30d})]. From this relation it follows that $\gradv\cdot \v E=\alpha\rho$ is not invariant unless we assume $\gradv \times\v B=\beta\v J$, which is invariant because $\gradv'=\gradv, \v B'=\v B$ and $\v J'=\v J$ in the magnetic limit. These results suggest that the Ampere-Maxwell law $\gradv\times \v B-(\beta/\alpha)\partial \v E/\partial t=\beta\v J$ must be replaced by $\gradv \times\v B=\beta\v J$. Hence, the magnetic limit of Maxwell's equations can be defined by
\begin{subequations}
\label{29}
\begin{align}
\gradv\cdot\v E&=\alpha\rho, \label{29a}\\
\gradv\times \v E+\frac{\alpha}{\beta c^2}\frac{\partial \v B}{\partial t}&= 0,\label{29b}\\
\gradv\cdot\v B&= 0 \label{29c}, \\ 
\gradv\times \v B&=\beta\v J.\label{29d}
\end{align}
\end{subequations}
The truncated Ampere-Maxwell law $\gradv \times\v B=\beta\v J$ should not be considered as the result of letting $\partial \v E/\partial t= 0$ in the complete Ampere-Maxwell law $\gradv\times \v B-(\beta/\alpha)\partial \v E/\partial t=\beta\v J$,
but as the result of assuming conditions that imply that $(\beta/\alpha)|\partial\v E/\partial t|$ is small with respect to $|\gradv\times \v B-\beta\v J|$.\cite{14} In Sec.~V we will discuss such conditions.

To verify the Galilean invariance of Eq.~(\ref{29}) we first use Eqs.~(\ref{21}), (\ref{27}) and (\ref{29}) and obtain
\begin{subequations}
\label{30}
\begin{align}
\gradv' \cdot \v E' - \alpha\rho'& = \gradv \cdot \bigg[\v E + \frac{\alpha}{\beta c^2}\bfv\times \v B\bigg]
 - \alpha\bigg[\rho - \frac{1}{c^2}\bfv \cdot \v J\bigg] \label{30a}\\
& = \gradv \cdot \v E - \alpha\rho + \frac{\alpha}{\beta c^2} \bigg[\gradv \cdot (\bfv \times\v B) + \beta\bfv \cdot \v J\bigg] \label{30b}\\
& = \gradv \cdot \v E - \alpha\rho - \frac{\alpha}{\beta c^2} \bigg[ \bfv\cdot(\gradv\times\v B) - \beta\bfv \cdot \v J\bigg] \label{30c}\\ 
& = \gradv \cdot \v E - \alpha\rho - \frac{\alpha}{\beta c^2} \bfv \cdot \bigg[\gradv \times\v B - \beta\v J\bigg] \label{30d}\\
& = \gradv \cdot \v E - \alpha\rho, \label{30e}\\
\gradv'\cdot\v B' &= \gradv\cdot\v B, \label{30f}\\
\gradv' \times \v E' + \frac{\alpha}{\beta c^2}\frac{\partial \v B'}{\partial t'}& = \gradv \times \bigg[\v E + \frac{\alpha}{\beta c^2}\bfv \times \v B\bigg] + \frac{\alpha}{\beta c^2}\bigg[\frac{\partial}{\partial t} + \bfv \cdot \gradv\bigg]\v B \label{30g}\\
 &= \gradv \times \v E + \frac{\alpha}{\beta c^2}\frac{\partial \v B}{\partial t} + \frac{\alpha}{\beta c^2}\bigg[ \gradv \times (\bfv \times \v B) + (\bfv \cdot \gradv)\v B\bigg]  \label{30h}\\
 &= \gradv \times \v E + \frac{\alpha}{\beta c^2}\frac{\partial \v B}{\partial t}, \label{30i}\\
\gradv' \times \v B' - \beta\v J'& = \gradv \times \v B - \beta\v J. \label{30j}
\end{align}
\end{subequations}
The Galilean invariance of Eq.~(\ref{29}) directly follows from Eq.~(\ref{30}).\cite{15}

The magnetic limit describes quasistatic situations where magnetic effects are dominant. The equations of the magnetic limit correspond to those of the quasimagnetostatics approximation,\cite{11} which describes capacitive but not inductive effects.\cite{12} Notice that the energy density in the magnetic limit $u=\chi\v B^2/(2\beta)$ depends only on the magnetic field. We note that magnetohydrodynamics relies on the magnetic limit.\cite{16} Equation~(\ref{29}) corresponds to ``electrodynamics before Maxwell."\cite{17}

The idea that Gaussian units are not suitable to derive the electric and magnetic limits is seen to be unfounded. If $\alpha=4\pi$ and $\beta=4\pi/c$ are inserted into Eqs.~(\ref{23}) and (\ref{29}), we obtain respectively the electric and magnetic limits of Maxwell's equations in Gaussian units.

\section{The instantaneous limit of Maxwell's equations}

The fact that the speeds $c_u$ and $c$ are distinct physical quantities allows us to violate the $c$ equivalence principle by changing $c$ and keeping $c_u$ unchanged. Such a formal possibility allows us to introduce the instantaneous limit of Maxwell's equations. By letting $c \to \infty$ in Eq.~(\ref{12}) we obtain
\begin{subequations}
\label{31}
\begin{align}
\gradv\cdot\v E&=\alpha\rho, \\
\gradv\times \v E& = 0,\\
\gradv\cdot\v B&= 0, \\
\gradv\times \v B-\frac{\beta}{\alpha}\frac{\partial \v E}{\partial t}&=\beta \v J,
\end{align}
\end{subequations}
which represent the instantaneous limit of Maxwell's equations. Equation~(\ref{23}) for the electric limit and Eq.~(\ref{31}) for the instantaneous limit have the same form,\cite{18} but the interpretation of these two sets of equations is different. The electric limit is a nonrelativistic limit because it is valid for $|\bfv| \ll c$. The instantaneous limit is not necessarily a nonrelativistic limit because it allows nonzero but otherwise arbitrary velocities.

To show the Galilean invariance of Eq.~(\ref{31}) we first derive the instantaneous limit of the relativistic transformations
\begin{subequations}
\label{32}
\begin{align}
\v E' &= \gamma\bigg[\v E + \frac{\alpha}{\beta c^2}\bfv \times \v B \bigg] + (\gamma - 1)\frac{\bfv(\bfv \cdot \v E)}{\bfv^2},\\
\v B' &= \gamma\bigg[\v B - \frac{\beta}{\alpha}\bfv \times \v E\bigg] + (\gamma - 1)\frac{\bfv(\bfv \cdot \v B)}{\bfv^2},\\
\rho' &= \gamma\bigg[\rho - \frac{\bfv}{c^2} \cdot \v J\bigg],\\
\v J' & = \v J - \gamma\bfv \rho + (\gamma - 1)\frac{\bfv(\bfv\cdot\v J)}{\bfv^2}.
\end{align}
\end{subequations}
By letting $c\to \infty$ (and then $\gamma=1$) we obtain the instantaneous limit of Eq.~(\ref{32}):
\begin{subequations}
\label{33}
\begin{align}
\v E'&=\v E, \label{33a} \\
\v B'&=\v B-\frac{\beta}{\alpha} \bfv\times \v E, \label{33b}\\
\rho'&=\rho, \label{33c} \\
 \v J' &= \v J -\bfv\rho. \label{33d}
\end{align}
\end{subequations}
Equations~(\ref{20}) and (\ref{33}) have the same form, but their interpretation is different. 
Equation~(\ref{20}) is valid when $|\bfv| \ll c,\;|\v E| \gg \alpha|\v B|/(\beta c)$ and $c|\rho| \gg |\v J|,$ while in Eq.~(\ref{33})
there is no restrictions on the magnitude of $\v E, \v B, \rho$ and $\v J$.

From the instantaneous limit in Eq.~(\ref{4}) we can obtain the Galilean transformations:
\begin{equation}
\gradv'=\gradv, \quad \frac{\partial}{\partial t'}=\frac{\partial}{\partial t}+\bfv\cdot\gradv, \label{34}
\end{equation}
which have the same form as Eq.~(\ref{21}). The interpretation of Eqs.~(\ref{21}) and (\ref{34}) is distinct because they were obtained from different equations [Eqs.~(\ref{2}) and (\ref{4})]. From Eqs.~(\ref{31}), (\ref{33}), and (\ref{34})] we can show the Galilean invariance of Eq.~(\ref{31}). This proof is formally the same as that given in Eq.~(\ref{24}) to show the Galilean invariance of Eq.~(\ref{23}) and so it is not necessary to show the proof here. Equation~(\ref{31}) has been discussed in SI units (without considering that they are the instantaneous limit of Maxwell's equations) in pedagogical papers\cite{12, 19} and in a research paper.\cite{20}

\section{Alternate approach to the electric and magnetic limits}

In Sec.~III we defined the electric and magnetic limits of Maxwell's equations following the indirect approach of Levy and Leblond.\cite{1} Can the electric and magnetic limits be directly obtained from Maxwell's equations? We next develop an alternative heuristic approach to answer this question in the affirmative.

{\it Alternative approach to the electric limit.} We have seen that the conditions $|\bfv|\ll c$, $|\v x| \ll ct$, and $|\v E|\gg\alpha|\v B|/(\beta c)$ imply $\gradv'=\gradv$ and $\v E'=\v E$. Hence
\begin{equation}
\label{35}
\gradv'\times \v E'=\gradv\times \v E.
\end{equation}
If only $|\bfv|\ll c$ and $|\v x| \ll ct$ are used, then $\gradv'=\gradv$ and $\v E'=\v E +[\alpha/(\beta c^2)]\v\bfv\times\v B$, and thus
\begin{equation}
\label{36}
\gradv'\times \v E'=\gradv\times \v E-\frac{\alpha}{\beta c^2}(\bfv\cdot\gradv)\v B,
\end{equation}
where we have used $\gradv\cdot\v B=0$. Equation~(\ref{36}) reduces to Eq.~(\ref{35}) if we impose the condition
\begin{equation}
\label{37}
|\gradv\times \v E| \gg \frac{\alpha}{\beta c^2}|(\bfv\cdot\gradv)\v B|,
\end{equation}
which is equivalent to the condition $|\v E|\gg\alpha|\v B|/(\beta c)$ in the sense that both conditions led to
Eq.~(\ref{35}). None of these conditions imply that $\gradv\times \v E=0$, and we require an additional assumption. If 
\begin{equation}
\label{38}
|(\bfv\cdot\gradv)\v B|\sim \bigg|\frac{\partial \v B}{\partial t}\bigg|,
\end{equation}
then Eq.~(\ref{37}) yields the condition
\begin{equation}
\label{39}
|\gradv\times \v E| \gg \frac{\alpha}{\beta c^2}\bigg|\frac{\partial \v B}{\partial t}\bigg|.
\end{equation}
When this condition is applied to Faraday's law, $\gradv\times\v E+ [\alpha/]\partial \v B/\partial t= 0$, we obtain $\gradv\times \v E=0$. The absence of the Faraday term in the electric limit should not be interpreted as the result of making $\partial\v B/\partial t=0$ in Faraday's law,\cite{9} but as the result of assuming conditions that imply that $[\alpha/(\beta c^2)]|\partial\v B/\partial t|$ is small with respect to $|\gradv\times \v E|$.

To elucidate the meaning of the condition in Eq.~(\ref{38}), we recall that $\partial \v B/\partial t$ represents the temporal variation of $\v B$ at a fixed point in space. In contrast, we can interpret $(\bfv\cdot\gradv)\v B$ as a convective derivative associated with the velocity $\bfv$ of the moving reference frame and with spatial variations of $\v B$. The condition in Eq.~(\ref{38}) implies that the local derivative of $\v B$ is of the same order as the convective derivative of $\v B$. That is, in the electric limit the values of the velocity $\bfv$ and of the spatial and temporal derivatives of $\v B$ are such that they satisfy Eq.~(\ref{38}).

We conclude that the condition that allows us to directly obtain the electric limit from Maxwell's equations [Eq.~(\ref{39})] arises from applying the conditions $|\bfv|\ll c$, $|\v x| \ll ct$, and $|(\bfv\cdot\gradv)\v B|\sim |\partial \v B/\partial t|$ to the relativistic transformation of the fields.

{\it Alternative approach to the magnetic limit}. From the conditions $|\bfv|\ll c$, $|\v x| \ll ct$, $|\v E|\ll\alpha|\v B|/(\beta c)$, and $c|\rho| \ll |\v J|$ we obtain $\gradv' = \gradv$, $\v B' = \v B$, and $\v J' = \v J$, and therefore
\begin{equation}
\label{40}
\gradv'\times \v B'-\beta\v J'=\gradv\times \v E-\beta\v J.
\end{equation}
If only $|\bfv| \ll c$ and $|\v x| \ll ct$ are used, then $\gradv' = \gradv$ $\v B' = \v B -(\beta/\alpha)\bfv \times \v E$, and $\v J' = \v J - \bfv\rho$. Hence
\begin{equation}
\label{41}
\gradv'\times \v B'-\beta\v J'=\gradv\times \v E-\beta\v J+\frac{\beta}{\alpha}(\bfv\cdot\gradv)\v E,
\end{equation}
where we have used $\gradv\cdot\v E=\alpha\rho$. Equation~(\ref{41}) reduces to Eq.~(\ref{40}) when the condition
\begin{equation}
\label{42}
|\gradv\times \v B-\beta\v J| \gg \frac{\beta}{\alpha}|(\bfv\cdot\gradv)\v E|,
\end{equation}
is imposed. The condition in Eq.~(\ref{42}) is equivalent to $|\v E|\ll\alpha|\v B|/(\beta c)$ in the sense that both conditions imply Eq.~(\ref{40}). None of these conditions guarantees the expected result $\gradv\times \v B-\beta\v J=0$. The additional required assumption is
\begin{equation}
\label{43}
|(\bfv\cdot\gradv)\v E|\sim \bigg|\frac{\partial \v E}{\partial t}\bigg|,
\end{equation}
which is used together with Eq.~(\ref{42}) to imply
\begin{equation}
\label{44}
|\gradv\times \v B-\beta\v J| \gg \frac{\beta}{\alpha}\bigg|\frac{\partial \v E}{\partial t}\bigg|.
\end{equation}
When this condition is applied to the Ampere-Maxwell law, $\gradv\times \v B-(\beta/\alpha)\partial \v E/\partial t=\beta\v J$,
we obtain $\gradv\times \v B=\beta\v J$. The absence of the displacement current in the magnetic limit should not be thought as the result of writing $\partial\v E/\partial t= 0$ in the Ampere-Maxwell law,\cite{14} but as the result of assuming conditions that imply that $(\beta/\alpha)|\partial\v E/\partial t|$ is small with respect to $|\gradv\times \v B-\beta\v J|$. Equation~(\ref{43}) states that the partial derivative of $\v E$ is of the same order as the convective derivative of $\v E$. In the magnetic limit the values of $\bfv$ and the spatial and temporal derivatives of $\v E$ are such that they satisfy Eq.~(\ref{43}).

We conclude that the condition that allows us to directly obtain the magnetic limit from Maxwell's equations [Eq.~(\ref{44})] arises by applying the conditions $|\bfv|\ll c$, $|\v x| \ll ct$, and $|(\bfv\cdot\gradv)\v E|\sim |\partial \v E/\partial t|$ to the relativistic transformation of the fields.

\section{The Lorentz force}

In the preceding sections we have considered the relativistic transformations of the fields to obtain their Galilean forms associated with the electric and magnetic limits of Maxwell's equations. This way of obtaining the electric and magnetic limits might have been developed  by a 
relativistic physicist. Let us imagine how a Newtonian physicist\cite{21} might have proceeded to obtain the Galilean transformations of the fields without the aid of special relativity.

Our Newtonian physicist has experimentally obtained the Lorentz force $\v F$ on a charge $e$ moving with the velocity $\v v$ in a region in which $\v E$ and $\v B$ fields are present:
\begin{equation}
\label{45}
\v F=e\bigg(\v E+\frac{\alpha}{\beta c_u^2}\v v\times\v B\bigg).
\end{equation}
This physicist is convinced that Eq.~(\ref{45}) is invariant under Galilean transformations because, according to Newton's second law, this force determines the acceleration which is Galilei invariant. By assuming that the charge $e$ acted on by the $\v E$ and $\v B$ fields is independent of the sources producing these fields and that the charge $e$ is Galilei invariant, the Newtonian physicist has no doubt of the validity of the equality
\begin{equation}
\label{46}
\v E'+\frac{\alpha}{\beta c_u^2} \v v'\times\v B'=\v E+\frac{\alpha}{\beta c_u^2}\v v\times\v B.
\end{equation}
The Newtonian physicist also knows the Galilean transformation of velocities
\begin{equation}
\label{47}
\v v'=\v v-\bfv.
\end{equation}
From Eqs.~(\ref{46}) and (\ref{47}) it is concluded that\cite{21}
\begin{equation}
\label{48}
\v E'+\frac{\alpha}{\beta c_u^2}\v v\times(\v B'-\v B)=\v E+\frac{\alpha}{\beta c_u^2} \bfv\times\v B'.
\end{equation}
The Newtonian physicist then looks for the field transformations that satisfy Eq.~(\ref{48}) and finds that the only physically satisfactory solution of Eq.~(\ref{48}) is given by
\begin{subequations}
\label{49}
\begin{align}
\v E'&=\v E +\frac{\alpha}{\beta c_u^2}\bfv\times \v B,\label{49a} \\
\v B'&=\v B\label{49b}.
\end{align}
\end{subequations}
The field transformations of the magnetic limit have been obtained without assuming any restriction on the magnitudes of the vectors $\bfv,\;\v E$ and $\v B$. Because Eq.~(\ref{49}) constitutes the only satisfactory solution of Eq.~(\ref{48}), the Newtonian physicist concludes that the field transformations of the electric limit are inconsistent with the Lorentz force. If experiments had led to the Lorentz dual force $\v F=e(\v B-(\beta/\alpha) \v v\times\v E)$, then our Newtonian physicist could have obtained the field transformations of the electric limit: $\v E'=\v E$ and $\v B'=\v B-(\beta/\alpha)\bfv\times \v E$ by requiring Galilean invariance of this unobserved dual force.

\section{Summary}
We have discussed three Galilean limits of Maxwell's equations. We pointed out that the Lorentz transformations admit three limits and noted the difference between the ultra-timelike and instantaneous limits of the Lorentz transformations. We emphasized the dual role of the speed $c$ in Maxwell's equations and pointed out that this role can be formalized by the $c$ equivalence principle. We have also shown that Maxwell's equations expressed in a form independent of specific units admit at least three Galilean limits. We emphasized the differences between the electric and instantaneous limits of Maxwell's equations. We also developed an alternative approach that allowed us to directly derive the electric and magnetic limits from Maxwell's equations. The electric limit of Maxwell's equations applies to quasistatic situations where the electric effects dominate the magnetic ones\cite{11,13} (the charge density dominates the current density) and the magnetic limit applies to quasistatic situations in which the magnetic effects dominate the electric ones\cite{11,16} (the current density dominates the charge density). The Lorentz force is consistent only with the magnetic limit of Maxwell's equations.

The present discussion on the Galilean limits can be used as complementary material in graduate and senior undergraduate courses on electromagnetism and special relativity. We suggest that a discussion of Maxwell's equations should include their Galilean limits. The Galilean world allows us to appreciate the results of the relativistic world and vice versa.

\begin{acknowledgments}

This paper is devoted to the memory of Oleg D. Jefimenko (1922--2009) who contributed changes to the conceptual understanding of electromagnetic theory.
\end{acknowledgments}

\section*{Table}

\begin{table}[h]
\centering
\begin{tabular}{|l|c|c|c|}
\hline
System & $ \alpha$ & $ \beta$ & $\chi$\\
\hline
Gaussian & $ 4\pi$ & $4\pi/c_u$ &$1/c_u$\\ \hline
SI & $1/\epsilon_0$&$ \mu_0$&$ 1$\\ \hline
HL & $ 1$ & $ 1/c_u$&$1/c_u$\\ \hline
\end{tabular}
\caption{\label{tab1}The $\alpha\beta\chi$-system of units.}
\end{table}

{}


\begin{thebibliography}{26}
\bibitem{5} Strictly, the ultra-timelike and ultra-spacelike limits are approximate limits. In contrast, the instantaneous limit is an exact limit. The ultra-timelike limit should be written as $\v x'\approx\v x- \bfv t$ and $ t'\approx t$ and the instantaneous limit as $\v x'=\v x- \bfv t$ and $t'=t$.
\bibitem{9}
If $\partial \v B/\partial t= 0$ is assumed, the electric limit becomes too restrictive because we would have magnetic fields constant in time, $\v B(\v x,t)=\v B(\v x,t_0)$, and electric fields linear in time, $\v E(\v x,t)=\v E(\v x,t_0)+(t-t_0)\dot{\v E}(\v x,t_0)$, where $t_0$ is a fixed time and the dot means time differentiation.
\bibitem{10}
Equation (23) implies the continuity equation $\gradv\cdot\v J+\partial\rho/\partial t = 0$, which is also Galilean invariant because of the relation $\gradv'\cdot\v J'+\partial\rho'/\partial t' = \gradv\cdot\v J+\partial\rho/\partial t$, where we have used $\gradv\cdot(\bfv\rho)=(\bfv\cdot\gradv)\rho$.
\bibitem{14}
If $\partial \v E/\partial t= 0$ is assumed in the magnetic limit, we would have electric fields constant in time, $\v E(\v x,t)=\v E(\v x,t_0)$, and magnetic fields linear in time, $\v B(\v x,t)=\v B(\v x,t_0)+(t-t_0)\dot{\v B}(\v x,t_0)$. The magnetic limit would be too restrictive.
\bibitem{15}Equation~(\ref{29d}) implies that $\gradv\cdot\v J=0$, which is Galilean invariant because $\gradv'\cdot\v J'=\gradv\cdot\v J$. Note that $\partial\rho/\partial t$ and $\gradv\times \v J$ are not related by the continuity equation in the magnetic limit.
Consider, for example, the charge density $\rho=-\gradv\cdot\v P$ and the current density $\v J= (1/\gamma)\gradv\times\v M$, where $\v P(\v x,t)$ and $\v M(\v x,t)$ are the polarization and magnetization vectors ($\v J$ does not include the polarization term $\partial\v P/\partial t$ because $|\v P| \ll \alpha|\v M|/(\beta c)$ in the magnetic limit). In this case we have $\partial\rho/\partial t\not=0$ and $\gradv\cdot\v J=0$.
\bibitem{18}
The electric and magnetic limits are approximate limits. The former should be written as $\gradv\cdot\v E\approx\alpha\rho$, $\gradv\cdot\v B\approx 0$, $\gradv\times \v E\approx 0$, and $\gradv\times \v B-(\beta/\alpha)\partial \v E/\partial t \approx\beta\v J$, and the latter as
$\gradv\cdot\v E\approx\alpha\rho$, $\gradv\cdot\v B\approx 0$, $\gradv\times \v E+ [\alpha/(\beta c^2)]\partial \v B/\partial t\approx 0$, and $\gradv\times \v B\approx\beta\v J$. In contrast, the instantaneous limit is exact: $\gradv\cdot\v E=\alpha\rho$, $\gradv\times \v E= 0$, $\gradv\cdot\v B=0$, and $\gradv\times \v B-(\beta/\alpha)\partial \v E/\partial t=\beta\v J$.
Thus there exists a subtle difference between the electric and instantaneous limits.
\end{thebibliography}
\end{document}